\documentstyle[]{article}
\begin{document}

\title{ Stochastic Diagrams for Critical Point Spectra}
\author{S.~Chaturvedi\thanks{Permanent address : School of Physics, 
University of Hyderabad, Hyderabad 500046 (India)}
~and  ~P.~D. Drummond\\
Department of Physics, 
University of
Queensland\\St Lucia 4067, Queensland, Australia.}
\date{\today } 
\maketitle

\begin{abstract}
A new technique for calculating the 
time-evolution, correlations and steady state spectra 
for  nonlinear stochastic differential equations is presented. 
To illustrate the method, we consider
examples  involving cubic nonlinearities 
in an N-dimensional phase-space. These serve as a useful 
paradigm for describing critical point 
phase transitions in numerous equilibrium and non-equilibrium systems. 
The technique presented here is not perturbative. It consists 
in developing the stochastic 
variable as a power series in time, and using this  to compute 
the short time expansion for the correlation functions. This, in turn, is 
extrapolated to large times
and Fourier transformed to obtain the spectrum.
 A stochastic diagram technique is developed to facilitate 
computation of the coefficients of the relevant power series expansion. 
Two different types of long-time extrapolation technique,
involving either simple exponentials or logarithmic rational 
approximations, are evaluated for third-order diagrams.
The analytical results thus obtained are compared with numerical 
simulations, 
together with exact results available in special cases.
The agreement is found to be excellent 
 up to and in the neighborhood of the critical point.
The exponential extrapolation works especially well
even above the critical point at large N-values,
where the dynamics is one of phase-diffusion
in the presence of a spontaneously broken symmetry.
 A  
feature of this method is that it also enables the calculation of the steady 
state spectra of polynomial functions of the stochastic variable. 
In these cases, the final correlations can be non-bistable even
above threshold, and the logarithmic rational
extrapolation has the greater accuracy.
Finally, we emphasize that the technique is also applicable
to more general stochastic problems involving spatial variation
in addition to temporal variation.

\end{abstract}


                                                                                                                                                                                                     
\section {Introduction} 

Stochastic differential equations are a natural way of describing
the interaction of a system with a random reservoir. They were introduced
by Langevin\cite{Langevin1908} to help explain Einstein's 
theory\cite{Einstein1905} 
of small particles immersed in a fluid, as observed by the
biologist Robert Brown.
More rigorous mathematical treatments were later
introduced by Ito\cite{Ito51} and Stratonovich\cite{Str66}.
They have now diffused into many different areas\cite{[1]} of physics, chemistry
and biology. In recent times, similar models have been utilized
in ever more diverse fields, including engineering, economics\cite{BlaSch73} ,
and even sociology. The essence of a stochastic differential
equation is that it isolates a system of interest from the
background of random events that may influence the system. Implicit
in this formulation is the idea that the reservoir, or source
of random fluctuations, evolves without reference to the system
of interest. This simplifies the study of  otherwise complex
coupled phenomena.

As an example,
the calculation of correlation functions, and hence the spectra,  of 
physical  systems near phase transitions is of  considerable theoretical  
interest. 
These have dynamical properties that are often conveniently
described using stochastic differential equations. However,
commonly used analytic techniques like linearization, 
frequently become invalid near 
phase-transition
 points. At  the same time, while numerical simulation is 
possible, this is a time-consuming computational procedure without  
resulting in a great deal of insight. Thus, there is a need for 
techniques that give analytical results. Rather surprisingly, there are
few systematic procedures for treating nonlinear stochastic
differential equations under conditions where linearization is invalid.

In  this paper we consider the spectra of physical systems that are 
described by stochastic differential equations near a critical point phase 
transition. Such differential equations have a near universal 
applicability, both for equilibrium and non-equilibrium phase transitions 
\cite{[1]}. For instance, they arise in theoretical treatments of single mode 
lasers \cite{[1]}, inhomogeneously and homogeneously broadened two mode  
lasers \cite{[2],[4]}, and optical parametric amplifiers near threshold 
\cite{[5],[6]}.
A number of useful theoretical techniques \cite{[7],[12]} are known for these 
problems, some of which improve upon the  
the conventional Zwanzig-Mori projection operator 
method \cite{[13],[14]}. 
However, these techniques are cumbersome for systems in higher 
phase space dimensions. 

Instead, we propose a simple, direct calculation which is based on the 
stochastic differential equation. The resulting integral 
expressions can be classified diagrammatically,  in a way that allows a 
straightforward calculation of essential combinatoric factors. The results 
give a power series in time which can be extrapolated to long times with 
reasonable accuracy in many cases.
 We analyze two possible extrapolation techniques,
namely the exponential of a rational function, and a series of simple
exponential terms. Either method gives excellent results at or below the
critical point. 

Above the critical point, we find differences in
accuracy, and this can be related to the dominant eigenvalue 
distributions for different types of equation and observable. Convergence
 is slowest when the spectrum has
characteristic time-scales
which are an exponential function of an equation parameter, as in 
the one-dimensional cubic stochastic process above threshold, which
involves diffusion over a barrier.
The exponential series method is preferable
for simple types of spectra with only one or two dominant eigenvalues,
which turns out to be the case for the $N$-dimensional
cubic stochastic process at large $N$-values. 
The rational function method is best for complex spectra  without much
range in characteristic time-scales. An example of this is the
intensity ($x^2$) correlation spectrum of the 
one-dimensional cubic process, which can
be represented with remarkable accuracy using low-order rational
function extrapolation.

More generally, we expect that this method can be applied to any
stochastic differential equation where conventional linearization methods
are inapplicable due to large nonlinear terms. Under these circumstances,
it may be useful to have a nonlinear solution of the type derived here,
as a starting point for a perturbative or asymptotic analysis. For these
reasons, it is useful to analyze the simple cubic nonlinear case in detail,
both as a test case for the stochastic diagram method, and as 
an elementary stochastic process of intrinsic interest.

\section{ Stochastic Equations}

The method of stochastic diagrams to  calculate solutions to stochastic 
differential equations is normally applied in the frequency domain, where 
it corresponds to a perturbation theory expansion in a small coupling  
constant \cite{[15],[16]}. In these applications, there is a close resemblance to 
Feynman 
diagram techniques. In both cases, the starting point of the iterative 
method is the approximate linearized solution to the problem, which 
becomes the solution to the entire correlation function in the limit as 
the coupling constant  approaches zero. Frequency domain stochastic 
diagrams have many useful applications, including the stochastic 
quantization approach to quantum field theory. An essential difference
between these methods and Feynman diagrams, is the appearance
of stochastic terms that are averaged over at a later stage.

   We choose here to apply stochastic diagrams to the time domain 
correlation function. This has the advantage that there are no 
singularities in the series expansion coefficients, even at a critical 
point. A corresponding disadvantage is that the long time correlation 
functions cannot be directly calculated, and must be approximated by an 
extrapolation procedure that is based on some known property of the 
solution. In the examples given here, we use 
either simple exponentials or logarithmic rational function 
extrapolation,
which results in analytic expressions
for the approximate correlation function.
 A direct comparison with numerically calculated spectra will be 
used to demonstrate the great 
accuracy of this procedure in calculating spectra 
near critical points. It is less accurate above threshold in the 
bistable cases where 
stochastic `barrier hopping' or `tunneling' can occur,
resulting in widely differing
eigenvalues. This  results in a reduced precision
for the extrapolation. Methods based on multiple time scales 
can be used  in these cases. 

Surprisingly,
in the closely related higher-dimensional phase-diffusion
problem,  stochastic diagrams  give good results.
An example of this is the laser above threshold. 
This is because the long time-scale here is not exponentially
long, as it is in the tunneling cases.
Good results are also found 
for non-bistable variables like the intensity, even when the underlying
equations are themselves bistable. 
This is because the extrapolation is carried out in terms of
the correlation function, which has a different behavior
to the underlying stochastic variable.
In order to illustrate these various
cases, we start with a very general form of stochastic
differential equation.

The equations we wish to treat  are of the form 
\begin{equation}
\dot{\bf x} = {\bf A}({\bf x}) +  {\bf B}({\bf x})\cdot \mbox{\boldmath $\xi$} 
(t)~~~~,
\end{equation}
where the real noise sources $\xi_i(t)$ have zero mean and are 
delta-correlated in time so that
\begin{equation}
<\xi_i(t)> = 0~~~~;
<\xi_i (t) \xi_j (t ^{\prime}) > = \delta_{ij}\delta(t- t^{\prime})~~~~~.
\end{equation}
Here ${\bf x}$ is a real vector of $n$-dimensions, ${\bf A}$ is an 
$n$-dimensional real polynomial vector function of ${\bf x}$ and ${\bf B}$ 
is an  $n \times m$ dimensional real polynomial matrix function of ${\bf x}$. 
The vector $\mbox{\boldmath $\xi$}(t)$ is an
m-dimensional real
 Gaussian stochastic process, interpreted in 
the It\^{o} sense\cite{[17]}, for computational simplicity.

\subsection{Iterative solutions}

The method of stochastic diagrams consists of performing an iterative 
solution for ${\bf x}^{(p)} (t)$ so that
\begin{equation}
{\bf x}^{(p)} (t) ={\bf x}_0 + \int_{t_0}^{t} dt^\prime [{\bf A}({\bf x}^{(n-1)} 
(t^\prime)) +{\bf B}({\bf x}^{(n-1)} (t^\prime))\cdot
\mbox{\boldmath $\xi$}(t^\prime)]~~~~~,
\end{equation}
where ${\bf x}^{(0)} (t)  \equiv{\bf x}_0\equiv{\bf x}(t_0)$.
 Next, correlation functions of the typical form
\begin{equation}
{\cal G}_{ij} (t,t_0) = <x_i (t) x_j (t_0)>- <x_i (t)>< x_j (t_0)>~~~~,
\end{equation}
are evaluated to $p^{th}$ order, resulting in an expansion of ${\cal 
G}_{ij} (t,t_0)$ as a power series in $\tau=t-t_0$ for $\tau > 0$. For any given 
term $p$ in the power series, iterations must be carried out until all possible 
terms in $\tau^p$ are evaluated. The result of the iterations consists of 
integrals over time which will be represented as directed lines in the 
stochastic diagrams. In addition, there are  polynomials in variables (represented as 
vertices), initial conditions in the variables
(represented as terminating arrows, and treated
as delta functions at the initial time) and stochastic 
terms (represented as crosses). 

\subsection{Cubic stochastic process}

Thus, for example, the solution of the well-known cubic 
stochastic process \cite{[17]}
\begin{equation}
\dot{x} = -x^3 +\xi(t)~~~~,
\end{equation}
has a first iteration, starting from a known initial
value $x(0)=v$ at $t=0$, of 
\begin{equation}
x^{(1)}(t) = v +w(t) - \int_{0}^{t} dt^\prime v^{3}~~~~.
\end{equation}
where $w(t) = \int_{0}^{t} dt^\prime \xi (t^\prime)$.
More generally, the $n^{th}$ iteration in this 
simple one-dimensional case is written
\begin{equation}
x^{(n)}(t) = v +w(t) - \int_{0}^{t} dt^\prime 
(x^{(n-1)}(t^\prime))^3~~~~.
\end{equation}
Thus, we can expand the second term in the iteration as 
\begin{eqnarray}
x^{(2)}(t) &=& v +w(t) - \int_{0}^{t} dt^\prime [v + w(t^\prime) 
-t^\prime v^{3}]^3
\nonumber \\ 
&=&v+w(t)  
+ \int_{0}^{t} dt^\prime [(t^\prime)
 ^3 v^{9} -3 (t^\prime)^2 v^{6} 
w(t^\prime) - 3 (t^\prime)^2 v^{7}-6t^\prime v^{4} 
w(t^\prime) \nonumber \\ 
&+& 3t^\prime v^{3} {w(t^\prime)}^2 + 
3t^\prime v^{5}-{w(t^\prime)}^3 - 3v^{2} w(t^\prime) -3v 
{w(t^\prime)}^2 -v^3]~~~~.
\end{eqnarray}
We see that even this simple example leads to a large number
of distinct terms, which need to be classified in a systematic way.
In particular,  while the leading term in the
integral is of order $t^4$, there are other terms
of lower order present, 
including the stochastic terms and  a term of order $t$ which
comes from the initial condition.

\subsection{Stochastic diagrams}

The next term in the iteration involves a cubic integral of $x^{(2)} (t)$, 
and clearly the combinatoric factors involved are more complex to three 
and higher orders. In order to simplify the counting of these factors, a 
diagrammatic classification can be introduced at this stage. In this 
classification, the terms are given diagrammatically to first order in 
Fig (1). To second order, all possible terms in $x^{(1)}(t)$ appear as 
`legs' on the nonlinear vertex, to the next higher order, as shown in Fig 
(2). 

Not all terms will contribute to the same order in a power series in $\tau$, 
since the expectation value of a product of two stochastic integrals is 
proportional to $\tau$, while the product of two deterministic integrals is 
proportional to $\tau^2$. This means that a reordering of the sequence is 
needed, to obtain a series of terms to a given order in $\tau$. The rules are 
simple: all vertices counts as one order in $\tau$ while stochastic terms 
count  as half an order in $\tau$ and initial values as zero order. If the 
terms in the reordered series are labeled as ${\tilde{x}}^{(n)}(t)$ we 
can represent them according to Fig (3).
On expanding all the relevant terms in Fig (3), we obtain:
\begin{eqnarray}
\widetilde{x}^{(0)}(t)&=&v\nonumber \\ 
\widetilde{x}^{(1/2)}(t)&=&w(t)  \nonumber \\ 
\widetilde{x}^{(1)}(t)&=& -v^{3}t\nonumber \\ 
\widetilde{x}^{(3/2)}(t)&=& -3v^{2}\widetilde{w}(t)\nonumber \\ 
\widetilde{x}^{(2)}(t)&=& 3[-v\widetilde{w^2}(t)+v^{5}t^2/2 ]\nonumber \\ 
\widetilde{x}^{(5/2)}(t)&=&-\widetilde{w^3}(t)+9v^{4}\int_0^t\widetilde{w}(t')dt'
+ 6v^{4}\int_0^tw(t')t'dt'
\nonumber \\ 
\widetilde{x}^{(3)}(t)&=&
3v^{3}\int_0^tw^2(t')t'dt' +9v^{3}\int_0^t\widetilde{w^2}(t')dt' 
\nonumber \\&+&18v^{3}\int_0^tw(t')\widetilde{w}(t')dt'  -t^3v^{7}-
{3 \over 2}t^3v^{7} \, .
\end{eqnarray}
Here we have introduced the notation of:
\begin{eqnarray}
\widetilde{w^n}(t)\equiv\int_0^t{w}^n(t')dt'  \, .
\end{eqnarray}
Further rules in stochastic calculus (of the It\^{o} variety) are that the 
expectation values of the products of initial terms with stochastic terms 
decorrelate to all orders at later times and all odd products of stochastic 
integrals average to 
zero. This means that the only surviving terms in the expectation value 
${\cal G}_{ij}$ must be 
the terms of integer order in the series. For other types of expectation
values (involving polynomials in $x(t)$), these extra terms must be retained.

If we take expectation values of the relevant stochastic terms, they
have the structure:
\begin{eqnarray}
\langle w^2(t)\rangle &=& t\nonumber \\ 
\langle \widetilde{w}(t)w(t)\rangle&=& 
\langle \widetilde{w^2}(t)\rangle=
t^2/2  \, .
\end{eqnarray}
Hence, on decorrelating, integrating over time, and combining all the
relevant terms up to third order we obtain 
(for the average and correlation function of $x(t)$):
\begin{eqnarray}
\langle x(t)\rangle &=&\langle v -t v^3+{3 \over 2}
t^2(v^5-v)+{1 \over 2}t^3(11v^3 - 5v^7) \rangle 
\nonumber \\ 
\langle x(t)x(0)\rangle &=&\langle v^2 -t v^4+{3 \over 2}
t^2(v^6-v^2)+{1 \over 2}t^3(11v^4 - 5v^8) \rangle  \, .
\end{eqnarray}
These results are valid for an arbitrary initial distribution function.
If carried out to higher orders,
it is clear that they can describe either a transient process, or
else a steady-state correlation in the time-domain, to any
order in time.

\section{N-dimensional cubic stochastic process}

Having introduced the stochastic diagram method, we now apply it to the 
N-dimensional cubic stochastic process
\begin{equation}
\dot{x_i}(t) =-\eta_i x_i -f_{ijkl}x_j x_k x_l +\xi_i (t)~~~;
~~i,j,k,l=1,\cdots,N~~~~~.
\end{equation}
Here summation over repeated indices is implied. The coefficient $f_{ijkl}$ 
of the cubic terms can be taken to be 
symmetric in the the last three indices without any loss of generality. 
This stochastic equation, with appropriate choice of parameters 
accommodates the stochastic equations that have been considered in the 
context of single and two mode lasers and optical parametric amplifiers 
[1-6].
The quantities of interest are the equilibrium correlation functions
\begin{equation}
{\cal G}_{ij}^{(n)} (\tau) = \lim_{t_0  \rightarrow \infty}
 [<x_i^n (\tau+t_0)x_j^n (t_0)>-<x_i^n(\tau+t_0)><x_j^n (t_0)>]~~~~.
\end{equation}
This equation now has the added feature of a linear loss/gain
term $\eta$. When $\eta>0$, there is additional damping, and
the system is below threshold
in the usual sense.
The deterministic critical point is at $\eta=0$.
When $\eta<0$, the system has linear gain (like a laser above the
lasing threshold), and the system is then above the critical point.
However, it is worth noting that as the dimensionality increases,
this type of classification which comes from a linearized
analysis is rather misleading. In fact, the 
enlarged phase-space volume means that noise sources become
increasingly important at large dimensionality - to the point that
there is a reduced distinction between the above and below threshold cases.

\subsection{Steady-state behavior}

Steady-state behavior is most readily analyzed if,
for simplicity, 
we  confine ourselves to the following equation with 
$N$-dimensional rotational symmetry: 
\begin{equation}
\dot{{\bf x}}(t) =-\eta {\bf x} -{\bf x} ({\bf x}\cdot {\bf x})/{N}  +
\mbox{\boldmath $\xi$} (t)~~~~.
\end{equation}
This corresponds to defining
 the cubic coefficient as:
$
f_{ijkl}=[\delta_{ij}\delta_{kl}+\delta_{ik}\delta_{jl}+\delta_{il}
\delta_{jk}]/(3N)~~~.
$

This stochastic equation has 
what is known as detailed balance - and hence
an exact solution in the steady-state, found by
examining the
corresponding Fokker-Planck equation:
\begin{equation}
{\partial \over \partial t}
P (t,{\bf x}) = {\bf L}_{FP}P (t,{\bf x})=
\sum_i\biggl[ \biggl({\partial \over \partial x_i}[\eta +
{\bf x} \cdot{\bf x}/N]x_i + {1 \over 2}
{\partial^2 \over \partial x_i^2}\biggr]P (t,{\bf x})
~~~.
\end{equation}

The equilibrium distribution is 
$
P_e ({\bf x}) = {\cal N} \exp{[-V({\bf x})]}$,
where $V( {\bf x})$ is a potential function given by:
\begin{equation}
V( {\bf x}) =
\eta {\bf x}\cdot {\bf x}+({\bf x}\cdot {\bf x})^2/2N
~~~.
\end{equation}

The stochastic
equation also has an exact relationship
between the moments of different orders in the steady-state,
which can be easily derived from the variable-change rules of Ito
stochastic calculus. These are:
\begin{equation}
{\cal M}^{(n+1)}
= ({2n+N-2}){\cal M}^{(n-1)}/({2N})
-\eta {\cal M}^{(n)}
\end{equation}

Here we have defined ${\cal M}^{(n)}=\langle [{\bf x} \cdot{\bf x}]^n \rangle_e/N$,
as a convenient normalized form of the moment. 
Although these recursion relations are useful, the
mean-square fluctuations have to be calculated
from the potential solutions given above. 
The quantity ${\cal M}^{(1)}=\langle{\bf x}\cdot {\bf x}\rangle_e/N$ 
can therefore be computed explicitly 
and is given by 
\begin{equation}
{\cal M}^{(1)}={\sqrt {\frac{N}{4}}}
{U\left(\frac{N+1}{2},{\sqrt N}\eta\right)}/{U\left(\frac{N-1}{2},
{\sqrt N}\eta\right)}~~~~,
\end{equation}
where $U(a,x)$ denote the Whittaker functions \cite{[18]}. 
For $\eta= 0$, this 
expression simplifies to 
\begin{equation}
{\cal M}^{(1)}={ \sqrt{\frac{2}{N}}}
{\Gamma\left(\frac{N+2}{4}\right)}/{\Gamma\left(\frac{N}{4}\right)}~~~~.
\end{equation}

\subsection{ $N=1$ case }

An important property of this potential 
in the one-dimensional case of $N=1$, is that it possesses
a potential barrier at $x=0$, 
if $\eta <0$. This means that 
there is a progression from a stable `below-threshold'
region for $\eta >0$, (where $x=0$ is the deterministic stable point),
to a critical region for $\eta =0$ characterized
by large fluctuations, and then to a bistable region 
for $\eta <0$. This is
characterized by local stability in the two potential wells
at $x=\pm \sqrt{|\eta|}$. 

The $N=1$ case has been well-studied in terms of its eigenvalue
spectrum. 
Any one-dimensional Fokker-Planck equation
 can be transformed to an equivalent
Schroedinger equation problem with imaginary times\cite{[7]}, 
by introducing a Schroedinger operator.
In this case, it has the
form:
\begin{eqnarray}
{\bf L}_{S}&=&
\exp{[V({\bf x})/2]}{\bf L}_{FP}\exp{[-V({\bf x})/2]}
\nonumber \\
&=& [2V''(x)-(V'(x))^2]/8+
{1 \over 2} {\partial^2 \over \partial x^2}
\nonumber \\
&=& -{1 \over 2}[x^3+\eta x]^2 +[\eta +3x^2]/2 +
{1 \over 2} {\partial^2 \over \partial x^2}
~~~.
\end{eqnarray}

At large positive values of $\eta$, the corresponding Schroedinger
potential reduces to a harmonic oscillator problem, with
quadratic potential. Transforming back to real time, the
eigenvalues of the Fokker-Planck operator are of form:
\begin{eqnarray}
{\bf L}_{FP}P_n ({\bf x})
&=& -\lambda_n P_n ({\bf x})
~~~,
\end{eqnarray}
where $\lambda_n = n\eta$. Physically this is easy to understand. The
equation is dominated by the linear decay rate $\eta$, and integer
multiples of $\eta$ will occur through the decay of integer powers
of the variable $x$.

At large negative values of $\eta$, the equation is bistable, and there
are two principle eigenvalues. 
A fast equilibration inside each potential well occurs,
with an eigenvalue of $\lambda_f = 2|\eta|$ in the limit of
large $\eta$.
A slow decay also occurs through diffusion
over the barrier. Ignoring terms in $|\eta|$ of order (1) 
in the pre-factor, this gives a slow eigenvalue:
\begin{equation}
\ln[\lambda_s] \simeq -\Delta V+\ln[|\eta|]
 \simeq -\eta^4/2+ \ln[|\eta|] \,  .
\end{equation}

It is significant for the
stochastic diagram method,
that this eigenvalue is exponentially small
in the limit of large $\eta$. Thus, we cannot expect an accurate
estimate of the eigenvalue with any technique involving a finite
series
of terms in $\eta$, and any corresponding spectrum in which $\lambda_s$
is significant will not be able to be estimated with a finite expansion
in powers
of $\eta$.

\subsection{ Large $N$ case }

For $N > 1$, a similar progression 
from below to above the critical point
holds deterministically, except that
there is no bistable region. Instead, for $\eta <<0$, there is a region 
characterized by neutral stability in the subset of phase-space
where $|{\bf x}|\simeq\sqrt{|\eta|N}$. Thus, there is a continuum of
possible deterministically stable behaviour. This phenomenon
is sometimes called spontaneous symmetry breaking. 
To show this more clearly, consider the distribution $P_R(R)$, in the
variable $R=|{\bf x}|^2/N$. This has a steady-state
potential of  $V_R(R) =N[\eta R + R^2/2-(1/2 -1/N)\ln(R)]$.
As a result, for increasing $N$, the distribution in $R$ is
peaked more and more strongly near the
value $R_{\eta}=(\sqrt{2+\eta^2}-\eta)/2$.
In fact, due to the increase in phase-space volume as $R$ increases,
there is always an outward `entropic' force even when $\eta>0$. This
means that the stochastic equation at large $N$ is not
described well by the deterministic stability theory.

In this limit,  the radius 
approaches a fixed
value, due to the balance between the outward entropic force due to
increasing phase-space volume, and the inward force due to the nonlinearity.
Thus, the
moments ${\cal M}^{(n)}$ all factorise, and are given by:
 \begin{equation}
{\cal M}^{(n)}=([\sqrt{2+\eta^2} -\eta]/2)^n ~~~~~.
\end{equation}
The recursion relation for moments now simplifies, and it is
straightforward to verify that the above solution does satisfy the recursion
relation.

Generally, in a
stochastic equation, spontaneous symmetry breaking is
accompanied by a type of phase-diffusion, or tangential
diffusion in a surface of dimension $N-1$. 
Hence, the lack of  bistability for any $N$ greater than one
results in a dynamical behaviour in which diffusion
still occurs, but with a reduced dimensionality.
These
two types of above-threshold behaviour result in different
dynamical regimes for the 
resulting correlation functions and
spectra. In both cases, the
above threshold dynamics typically
involves more than different time-scale.
The radial relaxation to a stable point within a potential well
in the $R$-space equations generally
occurs much faster above threshold than
the tangential diffusion.

There are corresponding changes in the large-N dynamics, and
the physical explanation for this 
is interesting. In the limit of $N \rightarrow \infty $,
the fast
radial equilibration takes place in 
an approximately quadratic potential well at all values of $\eta$. 
It may be noted that the 
corresponding (It\^{o}) stochastic equation 
for radial equilibration
involves 
noise in a multiplicative way. In the general case, we find that: 
\begin{equation}
\dot{R}=-2\eta R -2R^2 +1 +2\sqrt{R \over N}\xi(t)~~~~,
\end{equation}
where $\langle \xi(t) \xi(t')\rangle = \delta(t-t')$.
 As well as having multiplicative noise,
this equation also shows why the stochastic equation trajectories
are confined to an increasingly small region in $R$-space, as $N$ increases.
This occurs because the relative size of the noise term in the radial
equation decreases as N increases. Thus, radial equilibration
takes place with a fast relaxation rate of 
\begin{equation}
\lambda_{f}=  4R_{\eta}-2\eta = 2\sqrt{2+\eta^2} ~~.
\end{equation}

In the tangential direction,
diffusion takes place on a hyper-spherical surface of fixed radius
defined by
 \begin{equation}
|{\bf x}| =\sqrt{NR_{\eta}}\ . 
\end{equation}
Suppose
we define a coordinate system so the diffusion 
starts 
with a radial coordinate of $x_1=\sqrt{NR_{\eta}}$ at time $t=0$.
 For small times,
the other (tangential) coordinates obey the diffusion equation, so that 
\begin{equation}
\langle x_j^2\rangle=t \ ,
\end{equation} 
for $j>1$. Since the radius is fixed
by the radial equation, it follows that this 
corresponds to angular diffusion. Projecting each
angular variable in turn onto a radius vector in a lower dimensional subspace
reduces the length of the resulting vector.
Finally, in the subspace of  one dimension
containing the original (starting) vector, we have:
\begin{equation}
\langle x_1(t)\rangle\simeq\sqrt{NR_{\eta}
[1-t/NR_{\eta}]^{(N-1)}}\simeq\sqrt{NR_{\eta}}
\exp{[-(1-1/N)t/2R_{\eta}]} \  .
\end{equation} 

This corresponds to a  
much slower tangential
relaxation rate of 
\begin{equation}
\lambda_s =(1-1/N)/(2R_{\eta})
\end{equation} 
in the large N limit. We note that this is not exponential in $\eta$,
unlike the one-dimensional case. Similar behaviour would
occur in the case of finite $N$ and large, negative $\eta$,
which is also dominated by the tangential diffusion caused by
spontaneous symmetry breaking. However, for the case of finite
N values, the slow eigenvalue must reduce to 
$\lambda_s =\eta$ in the limit of large enough positive $\eta$.

\section{N-dimensional stochastic diagrams}

The $N$-dimensional
 equation clearly has the same structure as the integral equation 
associated with the simple cubic process  considered in the previous 
section, except for the linear terms which would complicate the diagrams
if retained. Instead, we can simply define $y_i (t)=\exp(\eta_i(t-t_0))x_i(t)$,
which obeys a stochastic equation without a linear term.
This can then be iterated as previously.
The iterative solution  has the same diagrammatic structure 
as before. Using an initial estimate for ${ x}_i(t)$ of
${x}_i^{(0)}(t)=e^{-\eta_i(t-t_0)} { v_i}  $,
where the initial value is $\bf {v} ={\bf x}(t_0)$,
the basic iterative solution is given by:
\begin{eqnarray}
{{ x}}_i^{(n+1)}(t) 
&=&e^{-\eta_i(t-t_0)} {x}_i(t_0)\nonumber \\ &+&
\int_{t_0}^{t}dt^{\prime}e^{-\eta_i 
(t-t^\prime)}[
\xi_i(t^\prime) -f_{ijkl}
{x_j}^{(n)}(t'){ x}_k^{(n)}(t'){ x_l}^{(n)}(t')]
\  . \nonumber \\
\end{eqnarray}
It is easily checked that replacing the approximations ${{\bf x}}^{(n)}$,
${{\bf x}}^{(n-1)}$ by ${{\bf x}}$, leads to an exact integral
equation for ${{\bf x}}(t)$.

We can now identify successive iterations with
terms in the stochastic diagrams  for  vector quantities
 $x_i (t)$, where each vertex includes a term $-f_{ijkl}$, and
 each directed arrow corresponds to
 $\int\exp(-\eta_i(t-t'))..$.
 Thus, in evaluating the diagrams,
  each vector initial
 condition is replaced by $v_j\exp(-\eta_j (t-t_0))$,
 and the noise term $w(t)$ is replaced by: 
 \begin{eqnarray}
 w_i(t)=\int_{t_0}^{t}dt_{1}e^{-\eta_i (t-t_1)} \xi_i (t_1)  
\end{eqnarray}.
To order $\tau^3$, this can be calculated using the diagrams 
in Fig (3a)-(3d), by making the associations given in Fig (4). 
Using the diagrams in Fig (3), one can easily derive an expansion for 
$x_i(t)$ up to order $\tau^3$. The details of the resulting
stochastic integrals are straightforward, but
rather lengthy. 

These results are given in the Appendix
for the rotationally symmetric case.In the symmetric case, it is 
also clear that ${\cal G}^{(1)}_{ij}(\tau) =0$ if 
$i\neq j$ and that
\begin{eqnarray}
{\cal G}^{(1)}(\tau) &\equiv& {\cal G}_{ii}^{(1)}(\tau)
\nonumber \\ &=&\lim_{t_0  \rightarrow \infty}
 [<{\bf x}(\tau+t_0)\cdot {\bf x}(t_0)>-<{\bf x}(\tau+t_0)>\cdot <{\bf 
 x}(t_0+\tau)>]/N~~~~.\nonumber \\
\end{eqnarray}


\subsection{N-dimensional two-time correlation function}

Averaging the expression for $x_i (t)$ thus 
obtained over $\xi(t)$'s, expanding the exponential factors and keeping all
terms up to order $\tau^3$, one obtains the result
for the steady-state or equilibrium two-time correlation function. 
In this expression the two-time correlation
is given in terms of the initial one-time moments of the
stochastic process. We note that it is not essential, at this
stage, to use equilibrium moments. The same general result
occurs regardless of the initial condition, even for
the case of transient correlations calculated {\it without}
taking the steady-state limit. 

We will focus on the symmetric case here, and
simplify the following
expression (derived in the Appendix), by using
the definition of $ {\cal M}^{(n)} ={<R^n>_e}= {<[{\bf x}\cdot {\bf x}/{N}]^n>_e}$,
where  the subscript $e$ denotes an equilibrium average: 

\begin{eqnarray}
{\cal G}^{(1)}(\tau) &=&
{\cal M}^{(1)}-\tau\left[\eta
{\cal M}^{(1)}
+{\cal M}^{(2)}\right] 
\nonumber \\
&+&\tau^2\left[\left(\frac{\eta^2}{2}-\frac{N+2}{2N}\right){\cal M}^{(1)}
+2\eta {\cal M}^{(2)}+\frac{3}{2}
{\cal M}^{(3)} \right]
 \nonumber \\
&-&\tau^3 \left[\left(\frac{\eta^3}{6}-\eta 
\left(\frac{5N+10}{6N}\right)\right){\cal M}^{(1)} \right.
\nonumber \\ 
&+&\left.\left(\frac{13\eta^2}{6}-\frac{7N+26}{6N}\right)
{\cal M}^{(2)}
+ \frac{9\eta}{2}{\cal M}^{(3)}
+\frac{5}{2}{\cal M}^{(4)}\right]  
\nonumber \\
&+& O(\tau^4)~~~.
\end{eqnarray}

Next, we can substitute the known equilibrium moments to obtain a
final expression for the correlation function
in terms of the mean square fluctuation  
${\cal M}^{(1)}=<{\bf x}\cdot {\bf x}/N>_e$, 
although still in a power series
in $\tau$. 
Using the previous relations
${\cal M}^{(n+1)}
= ({2n+N-2}){\cal M}^{(n-1)}/({2N})
-\eta {\cal M}^{(n)}
$, and defining 
\begin{equation}
a =1/(2{\cal M}^{(1)})\, ,
\end{equation}
one obtains the following power series:
\begin{eqnarray}
{\cal G}^{(1)}(\tau)&=&{\cal M}^{(1)}
\left[1- a\tau +\tau^2
\left(\frac{N+2}{4N}+\frac{1}{2}\eta a\right)
\right.\nonumber \\
&-&\left.\tau^3\left(\frac{N+8}{12N}a +
\frac{1}{6}\eta^2 a -\left(\frac{4-N}{12N}\right)\eta \right)
 + O(\tau^4)\right]~~~,
\end{eqnarray}
It is convenient to re-express this as:
\begin{equation}
{\cal G}^{(1)}(\tau)={\cal M}^{(1)}[1 -\sum_{n=1}^3 a_n\tau^n]
~~~.
\end{equation}
where:
\begin{eqnarray}
a_1 &=& a\nonumber \\
a_2  &=&-\left({N+2}+{2N}\eta a\right)/(4N)
\nonumber \\
a_3  &=&\left(({N+8})a +
2N\eta^2 a -({4-N})\eta \right)/(12N)
\end{eqnarray}
Although the result is still expressed in terms of the 
correlation function through $a$, this quantity
 can be calculated exactly, either by integrating the distribution
function numerically, or by using the Whittaker function representation. In the
case of $\eta= 0$, this  reduces to
$
a={\sqrt {N/8}}
{\Gamma\left({N/4}\right)}/{\Gamma\left(({N+2})/{4}\right)}$.

\subsection{Correlations of polynomials}

The diagrammatic expression for $x_i(t)$ in powers of $\tau$ 
and $W(\tau)$ can 
also be used to calculate 
the equilibrium correlations for any polynomial functions of $x_i(t)$. Thus,
 in the case $N=1$, for the equilibrium correlations for $R=x^2$:
\begin{equation}
{\cal G}^{(2)e} (\tau) = \lim_{t_0  \rightarrow \infty}
 <R(\tau+t_0) R(t_0)>-<R(\tau+t_0)><R(t_0)>~~~~,
\end{equation}
we obtain
\begin{eqnarray}
{\cal G}^{(2)e} (\tau) &=&[{\cal M}^{(2)}-({\cal M}^{(1)})^2] - 
\tau[2{\cal M}^{(1)}]+\tau^2[2-2\eta{\cal M}^{(1)}]
\nonumber \\
&-&
\tau^3[8{\cal M}^{(1)}+{4 \over 3}\eta^2{\cal M}^{(1)}]+O(\tau^4)
~~~.
\end{eqnarray}

Here we notice that ${\cal M}^{(2)} =1/2 - \eta{\cal M}^{(1)} $,
so the pre-factor in the above expression reduces to:
\begin{equation}[{\cal M}^{(2)}-({\cal M}^{(1)})^2] =
{\cal G}^{(2)}(0)={2a^2 - 2\eta a - 1 \over 4a^2}
~~~.
\end{equation}

Just as in the case above, we can write the
two-time correlation function
down in terms of
the individual power series terms, as:
\begin{eqnarray}
a^{(2)}_1 &=&4a/[2a^2 - 2\eta a - 1] \nonumber \\
a^{(2)}_2  &=&-[8a^2-4\eta a]/[2a^2 - 2\eta a - 1]
\nonumber \\
a^{(2)}_3  &=& 4a(4+2\eta^2/3)/[2a^2 - 2\eta a - 1]
\end{eqnarray}

\section{Spectral calculations}

The correlation function in the time domain must be extrapolated to 
long times in order to compute the spectrum, which involves a Fourier 
transform over all times. The general spectrum 
for any steady-state correlation function is:  
\begin{equation}
S^{(n)}(\omega) = 
2 Re \int_{0}^{\infty} d\tau {\cal G}^{(n)} (\tau) e^{i\omega 
\tau}~~~.
\end{equation}
In order to perform the Fourier transform, some extrapolation of the 
power series is required. 
In general, for an arbitrary initial condition, this is a
difficult operation to perform. However, in the steady state,
the un-subtracted correlation function must factorise at long times to the
product of the mean values at initial and final times. This means
that the correlation function defined here gives rise to
exponential decay at long times.

In fact, for the type of stochastic differential 
equations considered here, we expect a discrete spectrum with an 
exponential decay at long times. However, a simple truncation of the 
power series at a finite order will not lead to an exponential decay,
so we cannot just truncate the power series in time to obtain the spectrum.
We will consider two different approaches to extrapolation. The first
is to simply represent the correlation function with a finite series
of exponentially decaying terms,
the second is to approximate the logarithm of the correlation function
as a rational function.

We assume that our starting point is an arbitrary correlation function
${\cal G}$, expressed as power series 
up to $p$-th order in the stochastic diagrams,
of form:

\begin{equation}
{\cal G}(\tau)={\cal G}(0)[1 -\sum_{n=1}^p a_n\tau^n]
~~~.
\end{equation}

\subsection{Simple exponential extrapolation}

This technique represents the correlation function as a finite
series of
decaying exponential terms. The coefficients
can then be matched to the known
power series in time on a term-by term basis. This method is especially useful
when only a small number of eigenvalues dominates the spectrum. 

For a power series calculation to order $\tau^3$, 
two distinct exponential terms are required. More generally,
  any correlation function expanded
to order $p=2p'-1$ is represented using $p'$ effective eigenvalues as:    
\begin{equation}
{\cal G}(\tau)={\cal G}(0)\sum_{n=1}^{p'} g_n\exp(-\lambda_n\tau)
~~~.
\end{equation}
Here, for simplicity, we impose the restriction that $\sum_{n=1}^{p'} g_n=1$.
It is also obviously necessary that all the effective decay rates
are positive.
In the third order stochastic diagram case,
on matching powers of $\tau$, one obtains:

\begin{eqnarray}
g_1&=& {1 \over 2} -{a_1^3 + 3 a_1 a_2 + 3 a_3 \over 2 \Delta}         \nonumber \\
g_2&=&{1 \over 2} +{a_1^3 + 3 a_1 a_2 + 3 a_3 \over 2 \Delta} \nonumber \\
\lambda_1&=&{-\Delta -3a_3 -a_2 a_1 \over 2a_2+a_1^2}\nonumber \\
\lambda_2&=&{ \Delta-3a_3 -a_2 a_1 \over 2a_2+a_1^2}
\end{eqnarray}

where the denominator term $\Delta$ is defined by:
\begin{equation}
\Delta = \sqrt{6a_1^3a_3 -3a_1^2a_2^2 +18a_1a_2a_3-8a_2^3+9a_3^2}
~~~.
\end{equation}

\subsubsection{Amplitude correlations - N=1 case}

For the case of ${\cal G}^{(1)}$ 
at $N=1$, the dependence of $\lambda_1$ and $\lambda_2$ on 
$\eta$ is displayed in Fig (5a) by the  two solid lines, where the upper line 
corresponds to $\lambda_1$ and the lower line to $\lambda_2$. 
There is a marked transition in this $N=1$ case, between the non-bistable
behaviour for $\eta >>0$, where the two time-scales are identical,
and the bistable behaviour for $\eta <<0$, where one time-scale
becomes very long, corresponding to stochastic `barrier-hopping' over
the potential barrier in the
distribution function at $x=0$. In this
region, the extrapolation
technique used here is obviously less
reliable, since the relevant
eigenvalue is an exponential function
of $\eta$. One cannot accurately
estimate these long-time tails on the
correlation function, purely from the
short-time information provided by the
stochastic diagrams. In fact, there are
other techniques based on multiple
time-scales, which are more suitable in
this above-threshold region.

Nevertheless, the technique does generate 
the fast eigenvalue ($\lambda_f = 2|\eta|$)
correctly for large negative 
$\eta$. For large positive $\eta$, the harmonic oscillator predictions
are regained. It is interesting to note that the fast eigenvalue
in this case is $\lambda_f = 3|\eta|$; this occurs because the symmetry
of the problem means that even order eigenvalues are not significant
in the dynamics of $G^{(1)}(t)$ at large damping. The slow
eigenvalue is not accurately reproduced at large negative $\eta$,
since this becomes exponentially slow (i.e, should be a straight-line
graph).
We will show later,
by comparisons to numerical simulations,  that the critical dynamics are reproduced accurately.

\subsubsection{Amplitude correlations - N=4 case}

In Fig (5b), the behaviour of the eigenvalues for ${\cal G}^{(1)}$ 
at $N=4$ is shown
in the solid lines.
Here we expect the slow eigenvalue to approach $\lambda_s =.375/(R_{\eta})$,
where $R _{\eta}= [\sqrt{2+\eta^2} -\eta]/2$,
while the fast eigenvalue should be
 $\lambda_{f}=  2\sqrt{2+\eta^2}$. Since these are strictly large-N
 limits, we cannot expect these to be found exactly.
 These approximate
 results are actually reproduced with
 surprising precision, especially in the phase-diffusion limit
 of large negative $\eta$. Thus, we find $\lambda_s = .076 $
 and $\lambda_f = 10.4$
 at $\eta=-5$. The 
 approximate predicted values would be $\lambda_s = .074$,
 and $\lambda_f = 10.6$,
 with even better agreement at larger values of $|\eta|$.
  At large positive $\eta$, the 
 slow eigenvalue approaches $\eta$, and the fast
 eigenvalue approaches $3\eta$
 due to the $x^3$ term in the stochastic equation.
 This is  a result
 due (as in the $N=1$ case) to symmetry properties; the
 intra-well eigenvalue with $\lambda =2\eta$ does not contribute
 to this correlation function.
 
 \subsubsection{Intensity correlations - N=1 case}
 
 Finally, we consider a quadratic correlation function,
 which we term the intensity correlation. This is
 the case of $G^{(2)}$ at N=1. These results for the relaxation
 rates are given in the solid lines of Fig. (5c). Here we see no trace of
 the exponentially 
 slow eigenvalue. This describes a sign-reversal  process which has
 little or no effect on intensity correlations. Hence, all the
 observed relaxation rates are caused by the higher-order
 eigenvalues for intra-well relaxation. Far below threshold, 
 and above threshold,
 the eigenvalues
 approach $2|\eta|$ and $4|\eta|$,
 which are  characteristic
 of intra-well relaxation. Near the critical point
 of $\eta=0$, there is
 strong critical slowing down, with longest time-scales
 (smallest eigenvalue)
 being found at $\eta \simeq -1.5$. 
 Although this region is bistable, it
 gives
 the slowest  relaxation rate of
 $\lambda_s=2.35$; going further into the bistable region
 speeds up the intra-well relaxation.

\subsection{Rational logarithmic extrapolation}

An alternative `generic' technique, is to approximate the logarithm of the correlation 
function as a rational function,
with a numerator of one order larger than 
the denominator. This is guaranteed to have an exponential behaviour at 
large $\tau$. We call this procedure rational logarithmic
extrapolation. It can be applied to a power series of any order,
as long as it is known that the series gives rise to
exponential decay at long times. It is especially useful
for complex spectra that may have several eigenvalues - as long
as the values are not too different from each other.

For a power series calculation to order $\tau^3$, a quadratic rational 
function is required, so that we can approximate the correlation function 
to the given order as    
\begin{equation}
{\cal G} (\tau)={\cal G} (0)
\left[\exp-{\left(\frac{\alpha\tau+\beta\tau^2}{1+\gamma\tau}\right)}\right]~~~~.
\end{equation}
On matching powers of $\tau$
with the previous power-series expression, one obtains
the following general results:

\begin{eqnarray}
\alpha&=& a_1        \nonumber \\
\gamma&=&-{2(a_3+a_1 a_2 +a_1^3/3) \over a_1^2 + 2a_2}\nonumber \\
\beta&=&a_1 \gamma+a_1^2/2+a_3
\end{eqnarray}

These can be used to obtain an extrapolated correlation function
in any of the cases treated here. However, it is clearly important
to ensure that the asymptotic coefficient, $\beta/\gamma$, is
positive - otherwise no decay will occur.

\subsubsection{Amplitude correlations - N=1 case}

In the expression for $G^{(1)}$, the logarithmic expansion gives:
\begin{eqnarray}
\gamma&=&\frac{4Na^3-6N{\eta a^2} +2N{\eta^2 
a}-2{(N-1)a}-{(4-N)\eta}}{{3N+6}+
6N\eta a-6N{a^2}}~~~,
\nonumber \\
\frac{\beta}{\gamma} &=& a -
\frac{
\left(3N+6+6N(\eta 
a-a^2)\right)^2}{12N( 4N{a^3}-6N{\eta a^2} +
2N\eta^2 a-2(N-1)a-(4-N)\eta)}~.
\end{eqnarray}

The approximate expressions for the equilibrium correlations of $x(t)$ 
given above are characterized by two time scales - $\lambda_s = \alpha =a$ and 
$\lambda_l = \beta/\gamma$ which govern the short and long time behaviors 
respectively. For the case of ${\cal G}^{(1)}$ 
at $N=1$, the dependence of $\lambda_s$ 
and $\lambda_l$ on 
$\eta$ is displayed in Fig (5a) by the two dashed lines. 
As previously,
there is a marked transition in this $N=1$ case, between the non-bistable
behaviour for $\eta >>0$
and the bistable behaviour for $\eta <<0$.
Since the two time-scales here correspond to overall rates at long
and short times, and not effective eigenvalues, the distribution
of rates is different below threshold - eigenvalues with a low weighting do
not contribute very much to the final rate. For this reason, 
we see no direct evidence
for the fast time-scale below threshold, which corresponds to the
relaxation of higher order eigenstates.

 In the
region of long time-scale
`barrier-hopping', the extrapolation
technique used here is less
reliable. One cannot accurately
estimate these long-time tails on the
correlation function, purely from 
short-time information.
 This can be seen most clearly in the way
that the slower of the two time-scales goes off the bottom of the
logarithmic graph. At this stage, the longest time-scale is
negative, indicating that the rational function approximation
has broken down, and would predict an infinite or diverging
spectrum. Obviously, the extrapolation is severely
inaccurate at this point, and cannot be used this far above threshold.

\subsubsection{Amplitude correlations - N=4 case}

One might expect that the rational
approximations used here should improve
above threshold as $N$ increases, as the equations are
not bistable for large $N$.
This hypothesis proves to be valid, as
we show by the use of numerical
stochastic techniques in the following
sections. However, the improvement can already be seen in Fig (5b).
In Fig (5b), which gives ${\cal G}^{(1)}$ at $N=4$,
the slowest relaxation times above threshold are slightly too small
compared to the exponential method, which indicates that the
ratio of relaxation times is still too large for this method
to give correct extrapolations, although the situation is much
better than in the bistable case with $N=1$. 

\subsubsection{Intensity correlations - N=1 case}

For intrinsically non-bistable quantities like ${\cal G}^{(2)} (\tau)$ the
problems  above do not occur at all, as shown in Fig (5c),
which graphs ${\cal G}^{(2)}$ at $N=1$.  Here the results of both
extrapolation methods give similar behaviour. This
is not immediately evident from the graphs, as the rates
defined here do not have identical interpretations.
In fact, the generic method of rational function extrapolation
is actually better than the exponential method in this case.
 In order to demonstrate this
in detail,
we must turn to the full spectral calculation, which
include both the relaxation rates and the relative weights.

Before turning to the spectral results, we note that
the two-time correlation function for the
quadratic correlations in the rational approximation have quite
a simple form.
Extrapolating  to large $\tau$ using the 
rational function approximation as above, we obtain 
\begin{equation}
{\cal G}^{(2)} (\tau)\simeq[{\cal M}^{(2)}-[{\cal M}^{(1)}]^2] 
\left[\exp-{\left(\frac{a^\prime\tau+b^\prime\tau^2}{1+c^\prime\tau}\right)}
\right]~~~,
\end{equation}
where, using the result ${\cal M}^{(2)} =1/2 - \eta{\cal M}^{(1)} $,
\begin{eqnarray}
a^\prime &=& 4{\cal M}^{(1)}/(1-2\eta {\cal M}^{(1)} -2[{\cal M}^{(1)}]^{2})~~~,
\nonumber  \\
c^\prime &=& \left(\frac{a\prime^2}{3} 
+4+\frac{2}{3}\eta^2-\frac{1-\eta{\cal M}^{(1)}}{{\cal M}^{(1)}}a^\prime \right)/
\left(\frac{1-\eta{\cal M}^{(1)}}{{\cal M}^{(1)}}-\frac{a^\prime}{2}\right)~~~ 
\nonumber \\
\frac{b^\prime}{c^\prime}&=&a^\prime -a^\prime 
\left(\frac{1-\eta{\cal M}^{(1)}}{{\cal M}^{(1)}}-\frac{a^\prime}{2}\right)^2/
\left(\frac{a\prime^2}{3} 
+4+\frac{2}{3}\eta^2-\frac{1-\eta{\cal M}^{(1)}}{{\cal M}^{(1)}}a^\prime \right)~~~.
\end{eqnarray}

 It is interesting
to note  
from Fig(5c), that even for $N=1$, the $x^2$ variable is clearly not bistable,
and  shows no sign of the characteristic long time-scales of
bistable variables. Instead, there is a critical slowing-down near $\eta =0$,
with shorter time-scales at all other $\eta$ values. This also agrees
with the exponential extrapolation results.

\subsection{Spectral results}

Having computed a long-time extrapolation to the two-time
correlation function, it is now possible to calculate the spectrum. A 
simple Lorentzian spectrum is generated by  
the first order stochastic diagrams.
 This approximation we shall see is surprisingly 
close to the correct spectrum
even for finite $N$ values, especially at high frequencies. The reason 
for this is that the middle to high frequency spectral behaviour is 
mostly due to the change in slope in the two time correlation function 
near $\tau=0$, due to the fact that the steady state correlation function 
must be a function of $|\tau|$. The low frequency spectral behaviour near 
$\omega=0$ has additional contributions due to the $\tau \rightarrow 
\infty$ behaviour of the correlation function, which is not 
always given 
accurately using a first order expansion in $\tau$. The lowest 
order spectral contribution is therefore
\begin{equation}
S^{(1)}(\omega) = 2{\cal G}^{(1)} (0) Re \int_{0}^{\infty} d\tau
 e^{-a|\tau | +i\omega \tau} = \frac{1}{a^2 +\omega^2}~~~~,
\end{equation}
where $a=1/[2\langle R\rangle]$ is defined as in the previous sections.
Thus, in the large $N$ limit we expect to find that:
\begin{equation}
\lim_{N\rightarrow \infty}S^{(1)}(\omega) = 
 \frac{4R^2_{\eta}}{(1 +(2R_{\eta}\omega)^2}~~~~.
\end{equation}
To higher order, it is necessary to 
choose which extrapolation method to use.
 The  
procedure of extrapolating the correlation function to large 
$\tau$ by expanding it as a series of exponentially decaying terms,
is simple to Fourier transform, since this clearly results
in:
\begin{equation}
{\cal S}(\omega) = {\cal G}(0)\sum_{n=1}^{p'} 
{2{g}_n \lambda_n \over \lambda_n^2 +\omega^2} ~~~~.
\end{equation}
This   results in a  Fourier transform as a sum of 
Lorentzian components.  
For the three previous cases
 treated of $N=1$, $N=4$, 
 and the intensity spectrum for $N=1$,
 the spectrum thus obtained is plotted in Fig (6a) -(6c) 
for various values of $\eta$. The clear progression
from bistable behaviour, to spontaneous symmetry-breaking,
to non-bistable behaviour, is shown in these three graphs; as the
peak spectral intensity is greatly reduced for the spectra
with shorter characteristic time-scales. 

While this procedure has advantages as 
far as calculating the Fourier transform is concerned, it is not always the best
extrapolation. 
Accordingly, we also consider
a straightforward numerical Fourier transform, which can be used
 to calculate the 
spectrum from the rational function approximation to the correlation 
functions.

\subsection{Direct numerical simulation}

As there are no exactly known analytic expressions
for these spectra, we have to resort to stochastic numerical
techniques to check the accuracy of the spectrum using the
truncated diagram method. Thus,
in order to determine the accuracy of the correlation function, a direct 
simulation of the relevant stochastic differential equations was used. 
These simulations employed a strongly convergent semi-implicit numerical 
algorithm \cite{[19]}, with checks on both numerical sampling error and truncation 
error due to finite step-size. 
The actual algorithm used employed four iterations of the nonlinear
implicit equations at each step.
After starting the trajectories in a 
Gaussian distribution with variance equal to the known steady state 
variance, each trajectory was integrated for a total elapsed time much 
longer than the correlation time. 

Typically, $t=100-400$ was the maximum time 
used, with longer times being employed for calculations above threshold, 
where the correlation time is longer. In the simulations, the total 
number of trajectories employed was $10^6$, in order to 
reduce the sampling error to typically about $0.1-0.2\%$ at low frequencies -
although this typically rose to about $0.5\%$ above threshold,
presumably because of the highly non-Gaussian individual trajectory
statistics in these cases.
Sampling errors were
checked by subdividing the  results into $1000$
sub-ensembles, which were numerically Fourier transformed
and   averaged individually. The spectral results
were then averaged over the sub-ensembles, and the error-bar
 in the overall mean
was estimated using standard Gaussian distribution error estimates --
on the basis that sub-ensemble means should
have a Gaussian distribution according to the central limit theorem.

By 
reducing the time step to a small value (typically $\Delta t = 0.01 - 0.05 $), the 
errors due to the finite step-size were typically kept to below $0.5\%$.
This was estimated by calculating all spectra at two different step-sizes,
but with the same underlying stochastic noise terms, and comparing the 
results. The two error-bars were added to give the final numerical
error-estimates. No significant error from finite spectral windowing was found,
 although this would be expected to give problems in the
 extreme bistable regions.
 
By comparing with exact zero-frequency results, 
this technique
of error-estimation proved a reliable method, in the sense that
the discrepancies were of the expected size.

Thus, for example, the numerical spectrum 
for the case of $N=1$ at threshold ($\eta = 0$)
has the calculated value of $S(0) = 0.966  \pm    0.007$, using
$10^6$ trajectories, a window of $t_{MAX} = 200$, and a step-size  
of $\Delta t = 0.025 $. 
The error due to the finite step-size 
contributes $\pm    0.005 $ to the total error.
The corresponding exact result,
as explained in the next section, is $S(0) = 0.975$ - giving
a slightly greater actual numerical error than
the estimated one standard-deviation error-bar.
 By comparison, the extrapolated 
exponential series analytic
result is $S(0) = 0.970$, which is very close to the exact result.

For the case of $N=1$ above threshold ($\eta = -1.5$)
the calculated numerical simulation value is $S(0) = 10.01  \pm    0.10 $, using
$10^6$ trajectories, a window of $t_{MAX} =400$, and a step-size  
of $\Delta t = 0.05 $. 
Because the step-size is relatively large (in order to
maximize the time-window), the error-bar in this case
is mostly due to the finite step-size, which
contributes an error of $\pm    0.08 $ to the total error.
The corresponding exact result is $S(0) = 10.11$ - within
the estimated error-bars. By comparison, the extrapolated analytic
result is $S(0) = 9.06$, which gives an increased
extrapolation error, as expected.

The resulting numerical estimate of $S^{(n)}(\omega)$ was:
\begin{equation}
S_{num}^{(n)} (\omega) = 
\lim_{T\rightarrow \infty}<|\Delta{\widetilde{x_i^n}}(\omega)|^2>/T
~~,
\end{equation}
where the Fourier transform $\Delta{\widetilde{x_i^n}}(\omega)$ is defined as
\begin{equation}
\Delta{\widetilde{x_i^n}}(\omega)=
\int_{0}^{T} dt e^{i\omega t} [x_i^n(t)- \langle x_i^n(t) \rangle]  ~~~~.
\end{equation}

\subsection{Comparison of results}

The results of the two procedures, i.e. stochastic diagrams and numerical 
simulations, are compared in Figs (7), (8) 
and (9) for $N=1$, $N=4$ 
and the intensity spectrum with $N=1$ respectively. 
In each figure there are four lines, 
which are the Lorentzian approximation (dotted line), the 
rational approximation (dashed-dotted line), the
second order
exponential series extrapolation  (dashed line) and the direct numerical simulations 
(solid line).

In each figure there are sub-figures which correspond to 
different values of the driving field $\eta$, which are taken through a range of 
values from far below threshold to above
threshold. We notice that agreement is generally
excellent (close to the simulation
error-bars) 
for all methods
below threshold. Errors are always worst at low
frequencies, where the results are the
most sensitive to the 
extrapolation error at long times.
They are also  worst for the single
exponential extrapolation than the higher-order
extrapolation methods, as expected,
and usually best for the exponential series method.

\subsubsection{Amplitude correlations - N=1 case}

Below threshold, Fig (7a) shows the four different
spectral results near zero-frequency, thus
giving a magnified view of the results. Errors
are much smaller at higher frequencies, where all
the techniques agree to within the simulation
error-bars. It can be seen that
the exponential series method 
(dashed line) gives the
best low-frequency agreement 
to the simulation (solid line). The residual
difference is about equal
to the intrinsic numerical discretization and sampling errors,
while the other two methods give small,
but marginally significant discrepancies.

In Fig (7b), at the critical point of $\eta =0$, the
exponential series gives a
prediction at $N=1$ of $S(0) = 0.9702$.
This is in quite remarkable agreement with
the simulated
value of $S(0) = 0.966\pm 0.007$.
By comparison, the other two methods are again
either significantly higher (rational logarithmic),
or lower (Lorentzian approximation).

Above threshold, however, in Fig (7c), we see that the
agreement is outside the error-bars even for
the exponential series method, thus indicating a
reduced accuracy in the long-time extrapolation. This
is expected, in view of the fact that the
long-time eigenvalue is an exponential
function of $\eta$ - rather than a finite algebraic
expression, as would be generated from the stochastic diagrams.
Here the
errors increase to about $15\%$ for
$\eta = -1.5$, at zero frequency, 
in the rational approximation, with
much worse errors in the single
exponential approximation. 
However, in this case the exponential series still
gives the best result, with an error of less than $10\%$ .

\subsubsection{Amplitude correlations - N=4 case}

Fig (8) shows the spectrum of $x(t)$ as a function of $\eta$ 
at $N=4$.
In Fig (8a), with $\eta = 1$ (below threshold) the difference
between the various approximations and the
numerical simulation is about equal to the
sampling error-bars obtained with
$10^6$ trajectories. Thus, it is not possible to
distinguish between any of the extrapolations
in this case.
At threshold, in  Fig (8b), the differences increase, and the
exponential series method is clearly better
than a single exponential extrapolation.
Above threshold, in  Fig (8c), the accuracy
of both the single exponential and the rational 
extrapolation
diminishes,
relative to the exponential series method -
 due to the two dominant
eigenvalues
 in
this case. For
$N=4$, the above threshold error with the rational
extrapolation
reduced to $7\%$, as the multiple
time-scale problem is less significant
in this case. The error in the exponential
series method is less than $1\%$, i.e.,  of the same order
as the intrinsic sampling errors.

\subsubsection{Intensity spectrum - N=1 case}

Fig (9) shows the spectrum of $x^2(t)$ as a function of $\eta$ as given 
by the different approximations.  We see that there are 
obvious differences between this and the
previous cases. As $x^2(t)$ is 
not bistable, the spectrum does not have a large `spike' as $\eta 
\rightarrow -\infty$. This means that, unlike the previous examples,
the agreement between the rational function
extrapolation and simulation methods remains of the order of the numerical
sampling error-bars even above threshold. Since it would presumably
require more than $10^6$ trajectories to resolve the differences in these
spectral results, we have not attempted to accurately estimate
the extrapolation errors here. For this problem, it is clear that the
analytical theory is rather competitive with numerical simulations,
which are always subject to numerical sampling error. The nearly
perfect agreement for rational function extrapolation is 
a surprising result, given that it is obtained from only a small number
of stochastic diagrams. In this case, the exponential series method
gives slightly worse results, presumably because there are
several competing eigenvalues rather than just two.

\subsection{Exact results}

Although the agreement between the analytic results and the 
numerical simulations is generally excellent
(except for bistable variables) there are some features 
worth a closer examination. Firstly, we
emphasize again
 that discrepancies are only 
evident at low frequencies. This is simply because the correlation 
function tails, although only contributing a small part of the spectrum, 
cannot always be accurately extrapolated from the short time power series 
expansion. This discrepancy at or below the critical point is always 
small (of order of $1-2\%$ at the critical point $\eta=0$). The error is 
naturally larger in the simpler Lorentzian approximation, but in no case 
does it exceed $5\%$ with $\eta=0$. Secondly, the low frequency 
extrapolation error is larger above threshold as $\eta \rightarrow 
-\infty$, especially in the bistable case of $x(t)$ spectrum with $N=1$. 
This result is expected, since calculation
of the slow eigenvalue in this case 
requires an infinite series in $\eta$.

In the 
 particular case of $N=1$, the zero frequency spectrum is known exactly [7] and is given
  by 
\begin{equation}
S^{(n)}(0) = 4 \int_{-\infty}^{\infty} dx \frac{[f^{(n)}(x)]^2}{P_{eq}(x)}~~~~~,
\end{equation}
where
\begin{equation}
f^{(n)}(x) = -\int_{-\infty}^{x} dx_1  [x_1^n-<x^n>] P_{eq}(x_1)~~~~~.
\end{equation}
Here $P_{eq}(x)$ denotes the normalized equilibrium distribution. 
The discrepancy between this exact result at $N=1$, and the 
approximate spectra, 
is given in Fig (10). 

In the bistable case of $S^{(1)}$,
it is clear that the error increases 
very rapidly as $\eta \rightarrow -\infty$, especially for the single 
exponential and rational approximations. 
The reason for this is due to the well known bifurcation of the 
deterministic steady state in this case, when $\eta << 0$. The system can 
only switch from one solution near $x=\eta$ to the other near $x=-\eta$, 
on exponentially long time scales.
This can be seen on comparing Fig (11), which is computed at the
deterministic threshold of $\eta =0$, with
 an above threshold numerical 
simulation in Fig (12). In this region and above, a multiple time scale 
technique  would give 
the best results,
with other methods being used to estimate the slow eigenvalue.
 As we are interested here in the critical region around 
$\eta=0$, we simply note this problem here, rather than pursuing it in 
further detail. 
Techniques for dealing with multiple time-scales of these
barrier-hopping equations were first developed
by Kramers\cite{[20]}, and are quite well understood.

This type of problem 
is greatly reduced for $N>1$, where
there is a tangential diffusion above
threshold, rather than barrier-hopping.
This results in an increasing accuracy of the present
technique for large $N$, as
the system dynamics reduces to just two  time-scales,
both of which have finite expression in terms of $\eta$.
In this limit, the behaviour is closely analogous to the
well-known problem of diffusion in a curved space. While
there are no examples of physical systems with such large-dimensional
order-parameters,  many physical systems (like the laser above
threshold, or Bose-Einstein condensates) display this
type of spontaneous symmetry breaking accompanied
by tangential diffusion in phase-space. The case chosen ($N=4$)
is typified by a two mode laser with inhomogeneous broadening,
so there is no strong mode-competition. As we have seen, the stochastic
diagram method gives very accurate results when compared
with numerical simulations. However, we have no exact
result in this case.

Multiple time-scale problems
do not occur in the case of 
non-bistable quantities like $x^2(t)$ whose spectrum is given in Fig. 
(10b).
The exponential series 
method gives good agreement, as expected. 
Agreement with the rational
function extrapolation in this case is
excellent even above threshold. It is
so good that the exact spectral result at $\omega =0$ cannot
be told apart from the approximate value, so we have not
included the rational approximation in this graph.

\section{Summary}

The stochastic diagram technique is a method of classifying iterative 
terms in a stochastic power series expansion in time, so that terms of the same 
order in time are grouped together. This involves analyzing the 
deterministic and stochastic order, since they give rise to different 
types of contributions. 

In this paper we have used this technique to 
analyze critical amplitude and intensity spectra, by considering
either exponential series or logarithmic
rational function extrapolation. The results are inherently
 non-perturbative, and are very accurate except in bistable regions. 
Thus, the results are valid at the
critical point, where linearized spectra
diverge.
 Good results
are also obtained even above threshold
using exponential series extrapolation,
when there is
spontaneous symmetry-breaking.
The 
technique  works especially well for intrinsically non-bistable 
quantities like the intensity or radial spectrum. In these
cases, we find that rational function extrapolation is of
greater accuracy.

The general feature of these results 
for the case of the cubic stochastic process, is that we see the expected critical
slowing down at the threshold or critical point at $\eta=0$. Below threshold,
there is rather stable behaviour, except at high dimensionality where the
outward entropic effects of the phase-space volume are increasingly
important. Above threshold, there is a bistable region for $N=1$, where our
methods do not converge quickly, due to the exponentially long
time-scales involved. For this region, asymptotic methods involving
multiple time-scales would be more
suitable. For $N>1$,
our method  gives much better results than
for $N=1$ above threshold, since
the dynamics in this region are then dominated by a type of phase-diffusion
at fixed radius, rather than by barrier-hopping.

The problems treated here were of a relatively simple type,
to allow comparisons
with exactly known results.
We emphasize that the stochastic diagram
approach is rather general. It is not restricted to
equilibrium correlations, although extrapolation is not always
possible in transient calculations. Nor is it restricted to
problems just involving temporal variation: partial
stochastic differential equations can also be treated in this way.

In summary, this novel stochastic diagram technique appears to have
many possible applications for calculations involving stochastic
differential equations.
\\

One of us (SC) is grateful to the University of Queensland for the 
University of Queensland Travel Award for International Collaborative 
Research which made this work possible. 

\section*{ Appendix}
\setcounter{equation}{0}

In this appendix we briefly outline the details of the calculations 
leading to the result given in (34),
using the rules given in Fig (4). 

The contributions to the diagrams 
in Fig (3a) can easily be written down as follows
\begin{eqnarray} 
\widetilde{x_i}^{(0)}(t)&=&e^{-\eta t}v_{i} \nonumber \\
\widetilde{x_i}^{(1/2)}(t)&=&w_i(t)  
\end{eqnarray}
Next, evaluating the diagrams with one vertex in Fig (3b), gives:
\begin{eqnarray} 
\widetilde{x_i}^{(1)}(t)
&=&- f_{ijkl}\int_{0}^{t}dt_{1}e^{-\eta (t+2t_1)} 
v_{j}v_{k}v_{l} \nonumber  \\
\widetilde{x_i}^{(3/2)}(t)
&=& -3f_{ijkl}\int_{0}^{t}dt_{1}e^{-\eta (t+t_1)}
w_j(t_1)
v_{k} v_{l}
\end{eqnarray}
The next order terms in Fig (3c) include both one-vertex terms
with additional noise factors, and two-vertex nested integral terms:
\begin{eqnarray}
\widetilde{x_i}^{(2)}(t)
&=& 3f_{ijkl}\biggl[ -\int_{0}^{t}dt_{1}e^{-\eta t}
w_j(t_1)w_{k}(t_1)v_{l}\biggr.\nonumber \\ 
&+& \biggl. f_{lmnp}
\int_{0}^{t}dt_{1}e^{-\eta (t+2t_1)}  v_{j} v_{k}
\int_{0}^{t_1}dt_{2} e^{-2\eta t_2} v_{m} v_{n} 
v_{p} \biggr]
\nonumber\\
\widetilde{x_i}^{(5/2)}(t)
&=& 3f_{ijkl}\biggl[ -\int_{0}^{t}dt_{1}e^{-\eta (t-t_1)}w_{j}(t_1)
w_{k}(t_1)w_{l}(t_1) \biggr. \nonumber \\
&+&3   f_{lmnp}
\int_{0}^{t}dt_{1}e^{-\eta (t+2t_1)}  v_{j} v_{k}
\int_{0}^{t_1}dt_{2}
 w_{m}(t_2)
e^{-\eta t_2} v_{n}  v_{p}\nonumber \\ 
&+&2 \biggl. f_{lmnp}
\int_{0}^{t}dt_{1}e^{-\eta (t+t_1)}  v_{j}
w_{k} (t_1) \int_{0}^{t_1}dt_{3}
e^{-2\eta t_3} v_{m} v_{n} v_{p} \biggr]
\end{eqnarray}
Finally, the relevant third order terms are:
\begin{eqnarray}
\widetilde{x_i}^{(3)}(t)
&=&3f_{ijkl}e^{-\eta t} 
\biggl[ f_{lmnp}
\int_{0}^{t}dt_{1}
w_{j}(t_1) 
w_{k}(t_1)
\int_{0}^{t_1}dt_{2}
e^{-2\eta t_2} v_{m} v_{n}  v_{p} \biggr. \nonumber \\ 
&+& 3f_{lmnp}
\int_{0}^{t}dt_{1}e^{-2\eta t_1}  v_{j} v_{k}
\int_{0}^{t_1}dt_{2}
w_{m}(t_2)
w_{n}(t_2) v_{p} \nonumber \\ 
&+&6 f_{lmnp}
\int_{0}^{t}dt_{1}e^{-\eta t_1}  v_{j} 
w_{k} (t_1)
\int_{0}^{t_1}dt_{2}e^{-\eta t_2} w_{m}(t_2)
 v_{n}  v_{p}				\nonumber \\ 
&-&f_{kmnp}f_{lqrs}\int_{0}^{t}dt_{1}e^{-2\eta  t_1}
v_{j}
\int_{0}^{t_1}dt_{2}
e^{-2\eta t_2}v_{m}v_{n}v_{p}
\int_{0}^{t_1}dt_{3}e^{-2\eta t_3}v_{q}v_{r}v_{s}
						\nonumber \\ 
&-&3f_{lmnp}f_{pqrs}\int_{0}^{t}dt_{1}
v_{j}v_{k}
\int_{0}^{t_1}dt_{2}		
v_{m}v_{n}
\int_{0}^{t_2}dt_{3}e^{-2\eta (t_1 +t_2+t_3)}v_{q}v_{r}v_{s}\biggr] \, .\nonumber \\ 
\end{eqnarray}
In the above equations summation over repeated indices is implied. 
The next step consists in 

\noindent  
(a) multiplying the above expressions by $v_i=x_i(0)$ and summing over $i$ 

\noindent
(b) averaging the resulting expressions over the noise sources and the 
initial values and adding up all the contributions. 

\noindent
This leads to
 
\begin{eqnarray}
\frac{\langle{\bf x}(t)\cdot {\bf x}(0)\rangle }{N}&=&
 e^{-\eta t}\frac{1}{N}\langle {\bf v}\cdot {\bf v}\rangle 
 -\frac{1}{N} f_{ijkl} \langle v_{i} v_{j} 
 v_{k} v_{l}\rangle \int_{0}^{t} dt_1 e^{-\eta(t+2t_1)}
\nonumber \\
&+&\frac{3e^{-\eta t}f_{ijkl}}{N}\biggl[ -\delta_{jk} \langle v_{i} v_{l}\rangle
\int_{0}^{t} dt_1 
\int_{0}^{t_1} dt_2 e^{-2\eta(t_1- t_2)} \biggr. \nonumber \\
&+& f_{lmnp} \langle v_{i} v_{j}v_{k} v_{m}
v_{n} v_{p}\rangle
\int_{0}^{t} dt_1 
\int_{0}^{t_1} dt_2 e^{-2\eta(t_1+ t_2)}\nonumber \\
&+& f_{lmnp} \delta_{jk}\langle v_{i} v_{m}
v_{n} v_{p}\rangle
\int_{0}^{t} dt_1 \int_{0}^{t_1} dt_2
\int_{0}^{t_1} dt_3 e^{-2\eta(t_1-t_2+ t_3)}
\nonumber \\
&+&3 f_{lmnp}\delta_{mn} \langle v_{i} v_{j}v_{k} 
 v_{p}\rangle
\int_{0}^{t} dt_1 
\int_{0}^{t_1} dt_2 
\int_{0}^{t_2} dt_3 e^{-2\eta(t_1+t_2-t_3)}\nonumber \\
&+& 6 f_{lmnp}\delta_{km} \langle v_{i} v_{j}v_{n} 
 v_{p}\rangle
\int_{0}^{t} dt_1 
\int_{0}^{t_1} dt_3 
\int_{0}^{t_3} dt_2 e^{-\eta(t_1- t_2)}\nonumber \\
&-& f_{kmnp}f_{lqrs} \langle v_{i} v_{j}v_{m} 
 v_{n} v_{p} v_{q} v_{r} v_{s}\rangle
\int_{0}^{t} dt_1 \left(\int_{0}^{t_1} dt_2 e^{-\eta(t_1+2t_2)}
\right)^2\nonumber \\
&-& 3f_{lmnp}f_{pqrs} \langle v_{i} v_{j} v_{k} v_{m} 
 v_{n} v_{q} v_{r} v_{s}\rangle  \nonumber \\
&\times& \biggl.\int_{0}^{t} dt_1 \int_{0}^{t_1} dt_2 
\int_{0}^{t_2} dt_3 e^{-2\eta(t_1+t_2+ t_3)} \biggr] \, .
\end{eqnarray}
For 
the symmetric case with $f_{ijkl}$ given by $f_{ijkl}=[\delta_{ij}\delta_{kl}+
\delta_{ik}\delta_{jl}+\delta_{il}\delta_{jk}]/(3N)~~~$, 
the above results can be simplified. This is not essential to
the method, and neither is the use of a steady-state initial
condition at this stage.
On 
explicitly carrying out the summations, we obtain

\begin{eqnarray}
\frac{\langle {\bf x}(t)\cdot {\bf x}(0)\rangle}{N}&=&
\frac{e^{-\eta t}}{N} \biggl[ \langle {\bf v} \cdot {\bf v} \rangle  
 -\frac{1}{N} \langle ({\bf v}\cdot {\bf v})^2 
\rangle \int_{0}^{t} dt_1 e^{-2\eta t_1}\biggr.
\nonumber \\
&& -\frac{(N+2)}{N}\langle {\bf v}\cdot {\bf v}\rangle
\int_{0}^{t} dt_1 \int_{0}^{t_1} dt_2 e^{-2\eta(t_1-t_2)}\nonumber \\
&+& \frac{3}{N^2}\langle ({\bf v}\cdot {\bf v})^3\rangle
\int_{0}^{t} dt_1 
\int_{0}^{t_1} dt_2 e^{-2\eta(t_1+t_2)}\nonumber \\
&+&\frac{(N+2)}{N^2}\langle ({\bf v}\cdot {\bf v})^2 \rangle
\int_{0}^{t} dt_1 \int_{0}^{t_1} dt_2
\int_{0}^{t_1} dt_3 e^{-2\eta(t_1-t_2+t_3)}
\nonumber \\
&+& \frac{3(N+2)}{N^2} \langle ({\bf v}\cdot {\bf v})^2\rangle
\int_{0}^{t} dt_1 
\int_{0}^{t_1} dt_2 
\int_{0}^{t_2} dt_3 e^{-2\eta(t_1+t_2-t_3)}\nonumber \\
&+&\frac{2(N+8)}{N^2}\langle ({\bf v}\cdot {\bf v})^2\rangle
\int_{0}^{t} dt_1 
\int_{0}^{t_1} dt_3
\int_{0}^{t_3} dt_2 e^{ -\eta (t_1-t_2) }\nonumber \\
&-&\frac{3}{N^3}\langle ({\bf v}\cdot {\bf v})^4\rangle
\int_{0}^{t} dt_1 \left(\int_{0}^{t_1} 
dt_2 e^{-\eta(t_1+2t_2)}
\right)^2\nonumber \\ \biggl.
&-&\frac{9}{N^3} \langle ({\bf v}\cdot {\bf v})^4\rangle
\int_{0}^{t} dt_1 \int_{0}^{t_1} dt_2 
\int_{0}^{t_2} dt_3 e^{-2\eta (t_1+t_2+t_3)} \biggr] \, .
\end{eqnarray}
Finally, replacing the initial averages by equilibrium averages 
and expanding the time integrals in powers of $t$ we obtain
\begin{eqnarray}
\frac{\langle {\bf x}(t)\cdot {\bf x}(0)\rangle }{N}&=&
{\cal M}^{(1)}\left[1-\eta t +\frac{1}{2} \eta^2 t^2 -
\frac{1}{6}\eta^3 t^3 + \cdots \right]\nonumber \\
&-& {\cal M}^{(2)}\left[t -2 \eta t^2+ \frac{13}{6}\eta^2 t^3
+\cdots \right] 
\nonumber \\
&-&\frac{(N+2)}{N}{\cal M}^{(1)}\left[\frac{1}{2} t^2 -
\frac{5}{6}\eta t^3 + \cdots \right]\nonumber \\
&+&3{\cal M}^{(3)}\left[\frac{1}{2} t^2 -
\frac{3}{2}\eta t^3 + \cdots \right]\nonumber\\
&+&\frac{(N+2)}{N}{\cal M}^{(2)}\left[\frac{5}{6} t^3 + \cdots \right]
+\frac{2(N+8)}{N}{\cal M}^{(2)}\left[\frac{1}{6} t^3 + \cdots \right]\nonumber\\
&-&{\cal M}^{(4)}\left[\frac{5}{2} t^3 + \cdots \right] \, .
\end{eqnarray}
On collecting coefficients of like powers of $t$ we are led to the 
expression (34) for the short time expansion of the two-time correlation 
function in the N-dimensional case with rotational symmetry. This
also agrees with the earlier result of Eq (12), for the simpler case of $N=1$
and $\eta =0$ -- as expected.

\newpage
\begin{figure}
\caption{Diagrammatic representation of the iterative solution of the 
cubic stochastic equation to zeroth and first order.}
\end{figure}
\begin{figure}
\caption{Diagrammatic representation of the iterative solution of the 
cubic stochastic
 equation to second order.}
\end{figure}
\begin{figure}
\caption{Diagrammatic representation of the iterative solution of the 
cubic stochastic
 equation reordered by collecting together the diagrams which 
 contribute to the same power of $\tau$. The power of $\tau$ 
 that each diagram 
 contributes, is equal to the number of vertices plus 
 half the number of 
 crosses. Figs (a) -(d) indicate successively higher order
 diagrams, including fractional stochastic orders.}
\end{figure}
\begin{figure}
\caption{The rules for writing down the contribution of a given diagram.}
\end{figure}
\begin{figure}
\caption{The dependence of the apparent relaxation rates on 
the loss rate $\eta$ as given by the 
exponential series (solid line) and rational function 
(dotted line) extrapolations. Note that the two
types of extrapolation implicitly define different types of relaxation rate.
In each graph, the 
 lower of the two lines corresponds to 
 the slower relaxation rate, the upper to the faster relaxation rate.
 Cases treated are for 
(a) $ G^{(1)}$, $N=1$; (b) $ G^{(1)}$, $N=4$; (c) $ G^{(2)}$, $N=1$.}
\end{figure}
\begin{figure}
\caption{ The approximate equilibrium
spectrum for 
range of values of $\eta$, using the exponential series method.
Cases treated are for 
(a) $ S^{(1)}(\omega)$, $N=1$; (b) $ S^{(1)}(\omega)$, $N=4$;
 (c) $ S^{(2)}(\omega)$, $N=1$.}
\end{figure}
\begin{figure}
\caption{ Comparison of the equilibrium
spectrum $S^{(1)}(\omega)$ of $x(t)$ for $N=1 $ as 
given by stochastic numerical simulations (solid line), 
exponential series extrapolation
 (dashed line), 
rational function 
extrapolation (dashed-dotted line), 
Lorentzian approximation (dotted line) for 
(a) $\eta=1$, (b) $\eta=0$, (c) $\eta= -1.5$.
Error-bars for discretization and sampling
errors at zero frequency
are:
(a) $\Delta S=0.002$ , (b) $\Delta
S=0.007$ , (c) $\Delta S=0.1$.}
\end{figure}
\begin{figure}
\caption{ Comparison of the equilibrium spectrum 
$S^{(1)}(\omega)$ of $x(t)$ for $N=4 $ as
 given by stochastic numerical simulations (solid line), 
exponential series extrapolation
 (dashed line), 
rational function 
extrapolation (dashed-dotted line), 
Lorentzian approximation (dotted line) for 
(a) $\eta=1$, (b) $\eta=0$, (c) $\eta= -1.5$.
Sampling error-bars at zero frequency
are:
(a) $\Delta S=0.002$ , (b) $\Delta
S=0.01$ , (c) $\Delta S=0.1$.
}
\end{figure}

\begin{figure}
\caption{ Comparison of the equilibrium
spectrum $S^{(2)}(\omega)$ of $x^2 (t)$ for $N=1 $ as 
given by exact numerical simulations (solid line),
exponential series extrapolation
 (dashed line), 
rational function 
extrapolation (dashed-dotted line),  
 Lorentzian approximation (dotted line) for 
(a) $\eta=1$, (b) $\eta=0$, (c) $\eta= -1.5$.
Sampling error-bars at zero frequency
are:
(a) $\Delta S^{(2)}=0.0003$ , (b) $\Delta
S^{(2)}=0.001$ , (c) $\Delta S^{(2)}=0.004$.}
\end{figure}
\begin{figure}
\caption{ Comparison of the 
equilibrium spectrum  for $N=1 $ at 
zero frequency as 
given by the exact analytic expression  (solid line), 
exponential series extrapolation
 (dashed line), 
rational function 
extrapolation (dashed-dotted line),  
 Lorentzian approximation (dotted line) for 
a range of values of $\eta$.
Cases treated are:
(a) $S^{(1)}(0)$; (b) $S^{(2)}(0)$. }
\end{figure}
\begin{figure}
\caption{A sample numerical run showing critical fluctuations
 of $x(t)$ at 
threshold, for the case $\eta = 0$, $N=1$ .}
\end{figure} 
\begin{figure}
\caption{A sample numerical run showing bistability of $x(t)$ above 
threshold, for the case $\eta = -1.5$, $N=1$ .}
\end{figure}

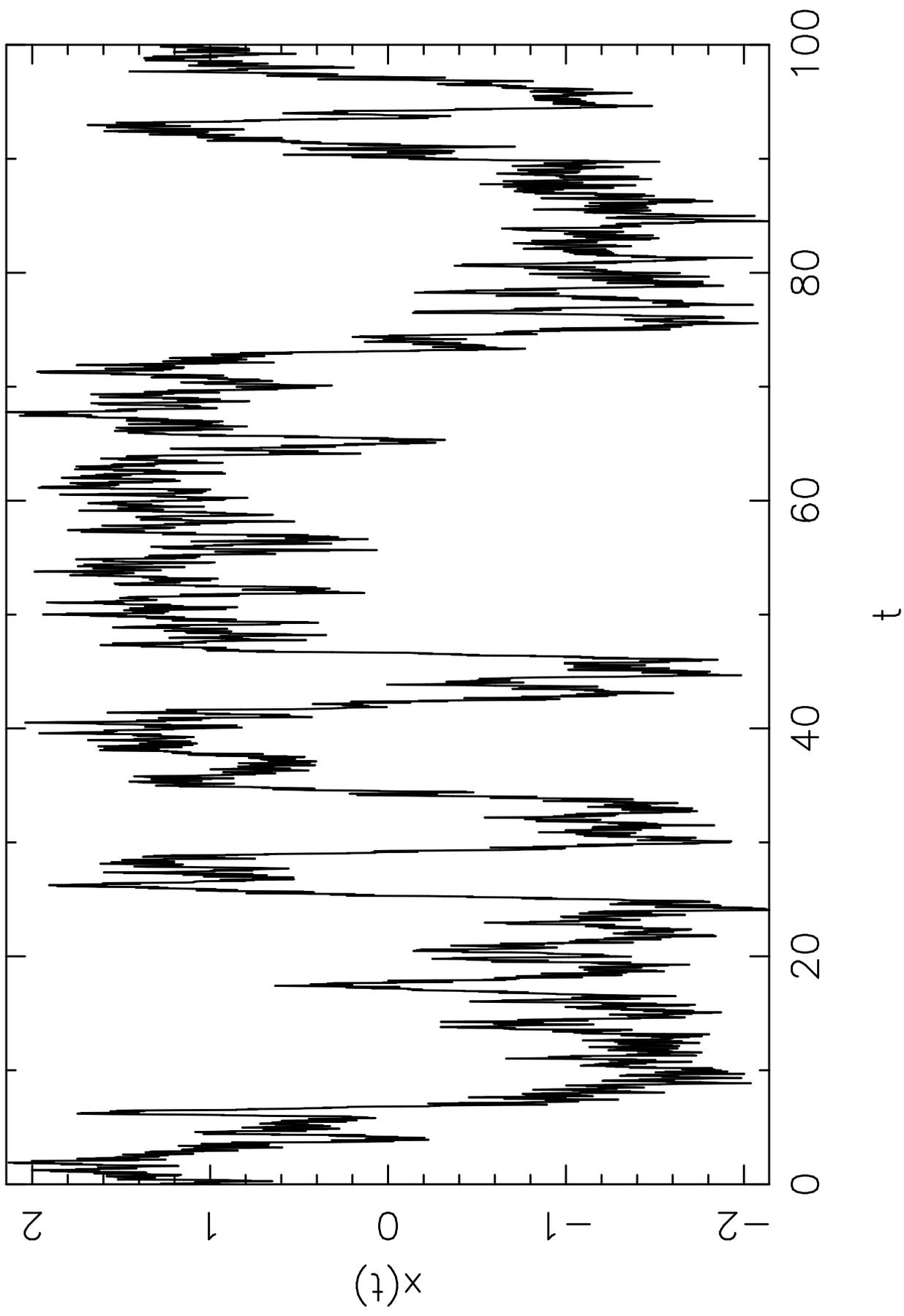
\end{document}